# Thermal-transport studies of Two-dimensional Quantum Spin Liquids

Minoru Yamashita, Takasada Shibauchi, and Yuji Matsuda

*Department of Physics, Graduate School of Science, Kyoto University, Kitashirakawa-oiwake, Sakyo, Kyoto 606-8502, Japan*

*Quantum spin liquids (QSLs) are fluid-like states of quantum spins where its long-range ordered state is destroyed by quantum fluctuations. The ground state of QSL and its exotic phenomena, which have been extensively discussed for decades, have yet to be identified. We employ thermal transport measurements on newly discovered QSL candidates, $\kappa$-(BEDT-TTF)$_2$Cu$_2$(CN)$_3$ and EtMe$_3$Sb[Pd(dmit)$_2$]$_2$, and report that the two organic insulators possess different QSLs characterized by different elementary excitations. In $\kappa$-(BEDT-TTF)$_2$Cu$_2$(CN)$_3$, heat transport is thermally activated at low temperatures, suggesting presence of a spin gap in this QSL. In stark contrast, in EtMe$_3$Sb[Pd(dmit)$_2$]$_2$, a sizable linear temperature dependence of thermal conductivity is clearly resolved in the zero-temperature limit, showing gapless excitation with a long mean free path (~1,000 lattice distances). Such a long mean free path demonstrates a novel feature of QSL as a quantum-condensed state with long-distance coherence.*

## Introduction

As the temperature is lowered, random thermal motion of particles becomes weaker, and usually a phase transition takes place when its ordering force overcomes the thermal fluctuation. The phase transition is accompanied by symmetry breakings and a long-range ordered state characterized by an order parameter sets in below the transition temperature. Phase transitions are ubiquitous from freezing of water to the very emergence of space-time in the early Universe. One of the most studied classes of such phase transition is magnetic ordering. In a paramagnetic phase, spins are free to rotate. Upon cooling below an ordering temperature, the rotational symmetry of spins is spontaneously broken and spins must choose a preferred direction. The ordered spins then produce a finite magnetization which characterises its long-range order as the order parameter. In addition to a class of such phase transitions driven by the thermal fluctuations, there is another class of phase transition driven by quantum fluctuations demanded by Heisenberg's uncertainty principle. The quantum fluctuations can persist down to the absolute zero temperature. Thus, a long-range order of spins can be destroyed by quantum fluctuations even at the zero temperature. The critical point where spins recover its rotational symmetry by quantum fluctuations is an explicit example of quantum critical point where an exotic state with a long-range quantum entanglement is expected to emerge.[1] The obtained paramagnetic state is, by analogy to an ordinary liquid, coined as a quantum spin liquid (QSL)[2] in which spins are highly correlated each other with a long distance due to the magnified quantum entanglement. The nature of QSL states has been attracting both theoretical and experimental attentions for decades. However, the detail description of the ground state or the elementary excitation which characterizes the QSL has remained elusive.

As the origin of quantum fluctuations of spins stems from the uncertainly principle, the smaller spins in lower dimensions feel the stronger fluctuations. In fact, in contrast to three-dimensional (3D) system where spins generally form a long-range ordered state, there is the celebrated Mermin-Wagner theorem that bans any long-range order of spins in one or two dimensions at a finite temperature when the Hamiltonian has a continuous symmetry.[3] Even at the absolute zero temperature, spins in one dimension are known to remain a disordered state. These ordered or disordered states can be characterized by different elementary excitations. For example, in 3D ordered spins, there are elementary excitations, called magnon, excited by flipping its spins. For quantum spins ($S$ = 1/2) coupling antiferromagnetically in 1D, the ground state is known to have an elementary excitation of domain walls as known as spinons. On contrary to these well established cases in one and three dimensional systems, understanding the ground state of quantum spins in two-dimension (2D) has remained an elusive issue, especially whether the ground state is a long-range ordered state or a QSL state at $T$ = 0. From theory, it has been considered that geometrical frustration plays a key role to realize a QSL. Geometrical frustration is a situation where none of spin configurations can minimize all the neighboring interactions simultaneously, e.g. Ising spins coupling antiferromagnetically on a triangular lattice (see Figure 1 (a)). In fact, a QSL state as known as resonating-valence-bond (RVB) state was first pointed out for quantum spins on 2D triangular lattice by P. W. Anderson at 1973.[4] This RVB state was lately revaluated as a possible mechanism for high-$T_c$ superconductors.[5] Since then, a lot of theoretical suggestions of exotic QSL states have been put force. These QSL states may possess exotic elementary excitations which obey either fermionic or bosonic statistics and have gapped or gapless energy dispersion.[2,6] Recently, promising candidates of QSL have been discovered in materials with 2D triangular lattice,[7][8][9] kagomé lattice[10] and pyrochlore lattice.[11] These progresses are stimulating further experimental and theoretical pursuits to reveal the ground state. Studies of QSL are now about to start flourishing.

Here, we summarize our thermal-transport studies done in the two organic compounds of QSL candidates, $\kappa$-(BEDT-TTF)$_2$Cu$_2$(CN)$_3$ [12] and EtMe$_3$Sb[Pd(dmit)$_2$]$_2$.[13][14] We find the two QSL states can be characterized by different elementary excitations; one is gapped and the other is gapless. We further find the gapless excitation has a very long mean free path which stretches ~1,000 lattice spacing, demonstrating its good quantum coherence as expected for a fluid-like state of quantum spins.

## Quantum spin liquids in two-dimension

Antiferromagnets consisting of 2D triangular lattices of Ising spins provide one of the prototypes of spin liquids under a geometrical frustration. As illustrated in Figure 1 (a), spins on at least one bond of each triangle must be chosen to be parallel. As a result, a large degeneracy of ground states remains even at the absolute zero temperature.[15] The spins do not form any long-range order but strongly correlate each other with a long correlation length. This geometrical frustration can be stronger in lattices where much more degenerate states are possible, such as in kagomé lattice (Fig.1 (b)). Ising spins are classical spins in materials where a strong axial anisotropy exists. On the other hand, the simplest model for quantum spins in isotropic spin space is the Heisenberg Hamiltonian

$$H = J \sum_{<i,j>} \vec{S}_i \cdot \vec{S}_j ,$$  Equation (1)

where $J > 0$ is a coupling constant for antiferromagnets, $\vec{S}_i$ and $\vec{S}_j$ are quantum spin operators for sites $i$ and $j$, respectively, and only the nearest-neighbor exchange is taken into account. In exchange for including quantum fluctuation, which can lower the energy of quantum spins, geometrical frustration becomes weaker in this model. In fact, in the case of the nearest Heisenberg quantum antiferromagnets ($S = 1/2$) in 2D triangular lattice, it has been shown that spins manage to order at $T = 0$ by orienting 120 degree each other (Fig.1 (c)).[16] Therefore, only the geometrical frustration is not enough to lead quantum spins to a QSL state on the triangular lattice.[17]

Recent progress of material synthesis, however, has reported promising candidates showing no sign of long-range ordered states at temperatures far below its interaction energy $J$. These include materials possessing 2D triangular lattice of quantum spins. For example, $^3$He atoms on graphite were found to form a 2D triangular lattice. Magnetization measurement down to 10 μK (~ $J$/300) has shown that the nuclear spins remain a quantum spin liquid without an energy gap.[7] Other recent candidates are organic compounds of κ-(BEDT-TTF)$_2$Cu$_2$(CN)$_3$ [8] and EtMe$_3$Sb[Pd(dmit)$_2$]$_2$ .[9] These organic compounds possess 2D layers of dimered molecules (BEDT-TTF for κ-(BEDT-TTF)$_2$Cu$_2$(CN)$_3$ and Pd(dmit)$_2$ for EtMe$_3$Sb[Pd(dmit)$_2$]$_2$, see Figure1 (d)). These dimers have one electron on each dimer, forming half-filling Mott insulating system with a 2D triangular lattice (Figure1 (e)). The coupling energy of spins ($J$) is estimated as ~250 K for both compounds from spin susceptibility measurements.[8][18] In spite of the large $J$, NMR measurements have shown that there is no internal magnetic field down to ~20 mK.[8][9] That is, the spins of these compounds remain a QSL state down to ~$J$/12,000. To our best knowledge, this is the lowest temperature compared with $J$ where a long-range ordered state is not found. The temperature of $J$/12,000 is even lower than temperature for 3D long-range order expected by a small inter-layer coupling ($T_{3D} \sim J$/100). Therefore, these spins remain a QSL state not because temperature is still high to observe its long-range-ordered ground state induced by the weak interlayer coupling, but because the system chooses the QSL state as the stable ground state.

These findings of QSL state at very low temperature indicate there is another mechanism stabilizing a QSL state, which is not included in the Heisenberg model. In fact, these materials locate near its insulating transition and the spins are expected to be influenced by a strong quantum fluctuation. Therefore, it is inappropriate for these spins to adopt the nearest-neighbour-only Heisenberg model which assumes that electrons are localized. To take into account further quantum fluctuation, theories with higher order perturbation have been proposed. An example is a multiple-spin exchange model where exchanges involving more than 2 spins are taken into account. From this model, it has been shown that the 4-spin exchange frustrates the 120-degree state and a QSL state may be stabilized when the 4-spin exchange is large enough.[19][20] Another successful strategy is to incorporate with the Hubbard model near the insulating transition. Detailed calculations have suggested that there is a QSL state in the insulating state close to the Mott transition.[21][22]

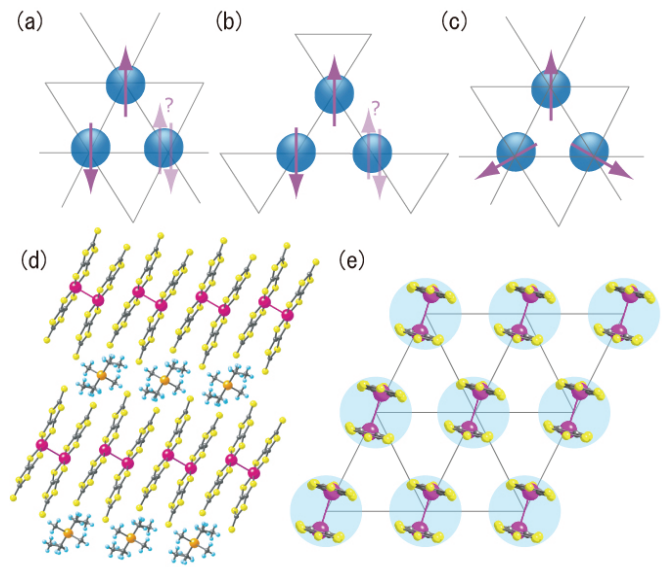

Figure1. (a, b) Illustration of Ising spins (arrows) coupling antiferromagnetically on a triangular lattice (a) and a kagomé lattice (b). (c) The 120 degree structure expected for the Heisenberg model on a triangular lattice. (d, e) The crystal structure of EtMe$_3$Sb[Pd(dmit)$_2$]$_2$; a view parallel (d) and perpendicular (e) to Pd(dmit)$_2$ layer.

## Experimental results and discussion

To identify a QSL state, knowledge about the low-lying excitations, especially about the presence or absence of an energy gap, is indispensable. For instance, in 1D antiferromagnetic spin chains, it is well known that elementary excitation for half-integer spin is gapless spinon,[23] whereas that for integer spin possesses the Haldane gap[24]. Probing these elementary excitations by experiments has played a pivotal role for understanding 1D spin chain systems. Thermal-transport measurement is a powerful tool to detect the elementary excitations down to very low temperatures. The advantage of this technique is the selective sensitivity to itinerant excitations carrying entropy, and is free from localized excitations by impurity effects (Schottky anomaly in specific heat, for example). We focus on the temperature dependence of thermal conductivity ($\kappa$), especially if there is a finite residual of linear-temperature dependence in the zero-temperature limit. This is often called $\gamma$ term, and the presence (absence) of the γ term immediately proves a gapless (gapped) excitation.

**EtMe$_3$Sb[Pd(dmit)$_2$]$_2$**

Here, we discuss thermal conductivity of EtMe$_3$Sb[Pd(dmit)$_2$]$_2$ and Et$_2$Me$_2$Sb[Pd(dmit)$_2$]$_2$.[13] Whereas these two insulators have a very similar lattice structure, the latter material exhibits a non-magnetic charge-ordered state below 70 K.[25] Therefore, the heat is only transferred by phonon ($\kappa = \kappa_{ph} \sim T^3$) in this non-magnetic compound, which enables us to separate the contribution of spins $\kappa_{sp}$ in the spin liquid material. At low temperatures, we confirm that $\kappa$ of Et$_2$Me$_2$Sb[Pd(dmit)$_2$]$_2$ shows the expected $T^3$ temperature dependence as shown in Figure 2. In stark contrast, we observe enhanced $\kappa$ in EtMe$_3$Sb[Pd(dmit)$_2$]$_2$ in the all temperature range we measured. Most remarkably, a finite γ term can be clearly resolved as $T \rightarrow 0$ K. Thermal conductivity is a product of heat capacity ($C_s$), velocity ($v_s$), and mean free path ($\ell_s$) of the elementary excitation. Since both $v_s$ and $\ell_s$ are nearly independent of temperature for $T \ll J$,[26] the finite residual of $\kappa_{sp}/T$ immediately means $C_s/T$ remains a finite value as $T \rightarrow 0$ K. Usually, a finite $C_s/T$ is associated to a property of normal metals where gapless electrons from the Fermi surface govern the heat capacity. This is therefore surprising result to find a γ term in EtMe$_3$Sb[Pd(dmit)$_2$]$_2$ because it is a completely *insulator* and electrons cannot be responsible for the observed γ. Hence, this finite $C_s/T$ in this insulating spin-liquid material leads us to conclude that there are gapless excitations of spins. This linear temperature dependence of heat capacity is also reported from recent heat capacity measurement.[27] By adopting the reported value of $C_s/T \sim 20$ mJ K$^{-2}$ mol$^{-1}$ and assuming $v_s \sim Ja/\hbar$ ($a \sim 1$ nm is the lattice constant), we can further estimate the mean free path as $\ell_s \sim 1$ μm, which means that the spin excitation can travel through 1,000 lattice distances without being scattered. Realizing such a long mean free path is a remarkable property of QSL in EtMe$_3$Sb[Pd(dmit)$_2$]$_2$, indicating the emergence of a quantum entanglement with a very long correlation length. We note that $\ell_s$ is much shorter for a spin transport in a paramagnetic state, a spin glass, or a VBC state. Moreover, the observed $\ell_s$ is even longer than that found in 1D spin chain materials[26] where a ballistic heat transport is proposed by theory.[28]

We further investigate the magnetic property of the elementary excitation by measuring the field dependence of $\kappa$ by applying fields perpendicular to the basal plane. We find $\kappa$ shows a flat field dependence below ~ 2 T followed by gradual increase at 0.23 K (Figure 3). This indicates that there are some magnetic excitations that couple to magnetic field, in addition to the above-mentioned gapless excitations responsible for the residual $\kappa_{sp}/T$ term at zero field. At higher temperatures ($T \geq 1$ K), the gapped field dependence evolves to a continuous increase. The overall field dependence is, therefore, understood by that there is additional magnetic excitations ($S \geq 1/2$) with an energy gap which opens below 1 K but closes for $H > 2$ T. Such an enhancement of thermal conductivity has been observed in a Haldane chain compound where additional excitation appears due to closing of a triplet-singlet gap under magnetic field.[29] The coexistence of the gapless excitations found in the zero field and the gapped excitation coupled with magnetic fields manifests another intriguing property of the QSL in EtMe$_3$Sb[Pd(dmit)$_2$]$_2$.

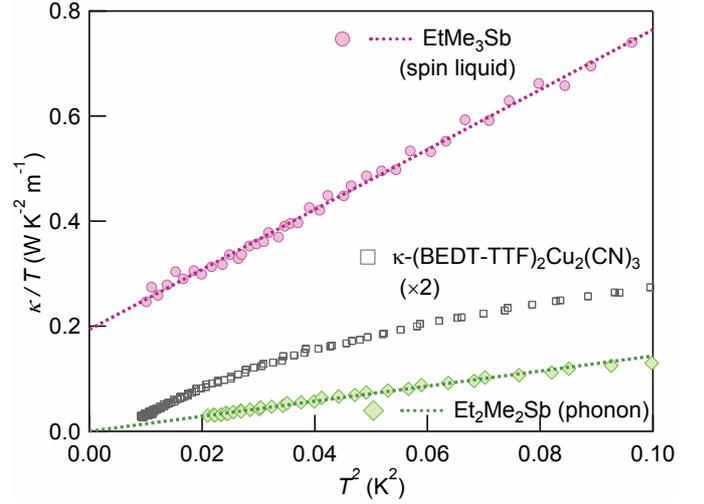

Figure 2. Thermal conductivity ($\kappa$) divided by temperature is plotted as a function of $T^2$ to see a residual of $\kappa/T$ as an intercept and a phonon contribution ($\kappa \sim T^3$) as a straight line. A clear residual of $\kappa/T$ of EtMe$_3$Sb[Pd(dmit)$_2$]$_2$ is resolved in the zero-temperature limit, whereas the non-magnetic compound Et$_2$Me$_2$Sb[Pd(dmit)$_2$]$_2$ with a similar lattice structure shows only a $T^3$ temperature dependence as expected from the phonon contribution. On the contrary, $\kappa/T$ of κ-(BEDT-TTF)$_2$Cu$_2$(CN)$_3$, which is multiplied by 2 for clarity, can be extrapolated to zero in the zero-temperature limit.

**κ-(BEDT-TTF)$_2$Cu$_2$(CN)$_3$**

The temperature dependence of $\kappa$ in κ-(BEDT-TTF)$_2$Cu$_2$(CN)$_3$ is convex at low temperatures[12] and its zero-temperature extrapolation goes to zero, which implies that there is an energy gap in the excitation spectrum. In fact, $\kappa$ at low temperatures can be well fitted by an activated behavior $\kappa \propto \exp(-\Delta/k_B T)$ with $\Delta \sim 0.46$ K. Such temperature dependence is quite different from the EtMe$_3$Sb[Pd(dmit)$_2$]$_2$ case, and indicates the absence of gapless excitaions that contribute to the thermal transport. The field dependence (Fig. 3) shows an increase of thermal conductivity for fields higher than ~ 4 T, suggesting a closing of a gap for a magnetic excitation as observed in EtMe$_3$Sb[Pd(dmit)$_2$]$_2$ at the lowest temperature. Very recently, these energy gaps have been consistently reported in μSR measurements.[30] The presence of the energy gap, however, is inconsistent with the heat capacity measurements reporting a finite γ term.[31] One interpretation for these observations is that the activated behavior of thermal conductivity is masked by the Schottky anomaly which plagues the heat capacity measurements at low temperatures.[31] An anternative explanation is that the discrepancy is originated from an inhomogeneity of QSL in κ-(BEDT-TTF)$_2$Cu$_2$(CN)$_3$. From longitudinal relaxation time ($T_1$) measurements,[32] it has been shown that NMR relaxation in κ-(BEDT-TTF)$_2$Cu$_2$(CN)$_3$ cannot be fitted to a single-exponential decay in low temperatures. Instead, they adopt a stretch-exponential form that requires a broad distribution of $T_1$.[33] In contrast, the NMR relaxation in EtMe$_3$Sb[Pd(dmit)$_2$]$_2$ shows a single-exponential decay below 1 K,[9] which implies a good homogeneity of the spin state at low temperatures. Therefore, on contrary to the gapless excitation in EtMe$_3$Sb[Pd(dmit)$_2$]$_2$ that can transport heat with a long mean free path due to the good homogeneity, the gapless spin excitation in κ-(BEDT-TTF)$_2$Cu$_2$(CN)$_3$ may be localized due to the inhomogeneity. Thermal-transport measurements are only sensitive to itinerant excitations, whereas heat capacity measurements detect all. Thus, the activated temperature dependence may be observed only in thermal transport measurements. Although this different degree of homogeneity provides an tentative explanation for the discrepancy, understanding of the different temperature dependence of $\kappa$ observed in the two QSLs should deserve further studies.

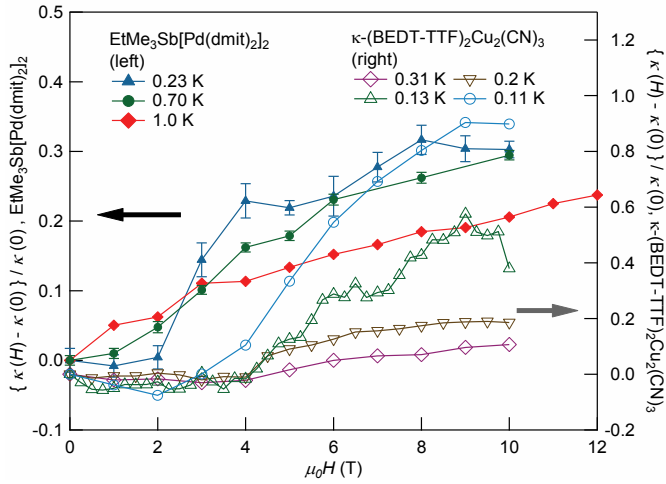

Figure 3. Field dependence of $\kappa$. The change under field normalized by the zero-field value is plotted.

So far, diverse theories of QSLs have been put forth in 2D frustrated spin systems. One intriguing candidate is a QSL with a Fermi surface of spinons.[34] This theory can naturally explain the finite $\gamma$ term observed by thermal conductivity and heat capacity measurements by analogy to electrons in normal metals. Moreover, the theory offers an unique oppotunity to identify the spinon Fermi surface by measuring a thermal Hall effect.[35] This thermal Hall effect is a heat transport version of the electronic Hall effect, in which elementary spin excitations curved by magnetic fields yield a thermal gradient perpendicular to the transport direction. We have tried to measure the thermal Hall effect in EtMe$_3$Sb[Pd(dmit)$_2$]$_2$ up to 12 T. However, within our resolution, any discernable thermal Hall effect has not been observed.[13] It remains a future work to check the spinon Fermi surface by other experimental techniques. On the other hand, the gapped excitation observed in $\kappa$-(BEDT-TTF)$_2$Cu$_2$(CN)$_3$ raises another interesting question about the minute gap size compared with $J$. It has been suggested that a topological vortex excitation in 2D QSL, called visons, may be responsible for the gapped excitation.[36] A recent $\mu$SR measurements[30] has aslo reported an energy gap of the similar size, together with another smaller gap. Identification of these gaps, however, remains an open question.

## Outlook

We report thermal-transport studies of two organic QSL candidates EtMe$_3$Sb[Pd(dmit)$_2$]$_2$ and $\kappa$-(BEDT-TTF)$_2$Cu$_2$(CN)$_3$. Most remarkably, in EtMe$_3$Sb[Pd(dmit)$_2$]$_2$, we find a linear-temperature dependence of $\kappa$ persists in the zero-temperature limit as evidence of a gapless spin excitation in this QSL. We further discover that the mean free path of the gapless reaches ~1,000 lattice distances. Such a nearly ballistic transport clearly demonstrates a strongly correlating spin state with a long correlation length. This is exactly the feature anticipated for a fluid-like state of quantum spins as originally coined as QSL. Moreover, a magnetic excitation with an energy gap of ~2 T is observed to coexist with the gapless excitation. On contrary, in $\kappa$-(BEDT-TTF)$_2$Cu$_2$(CN)$_3$, a tiny energy gap is inferred from the temperature dependence of $\kappa$. The contrasting behaviour found in the two organic materials would be only a small part of diverse properties of frustrated quantum spins. Further theoretical and experimental studies, including identification and comparison of the two QSLs, will reveal exotic properties of quantum spins as a new condensed state of matter.

## Acknowledgements

These studies have been done in collaboration with S. Fujimoto, R. Kato, M. Nagata, N. Nakata, T. Sasaki, Y. Sensyu, and D. Watanabe. We would like to thank M.A.P. Fisher, T. Hotta, M. Imada, K. Kanoda, H. Katsura, N. Kawakami, H. Kawamura, P.A. Lee, G. Misguich, N. Nagaosa, K. Totsuka and S. Watanabe for valuable discussions. The work was supported by KAKENHI from JSPS and a Grant-in-Aid for the Global COE Program `The Next Generation of Physics, Spun from Universality and Emergence' from MEXT, Japan.